# Specific Heat and Electrical Transport Properties of $Sn_{0.8}Ag_{0.2}Te$ Superconductor

Yoshikazu Mizuguchi[1]*, Akira Yamada[2], Ryuji Higashinaka[2], Tatsuma D. Matsuda[2], Yuji Aoki[2], Osuke Miura[1], and Masanori Nagao[3]

[1]*Department of Electrical and Electronic Engineering, Tokyo Metropolitan University, Hachioji 192-0397, Japan*
[2]*Department of Physics, Tokyo Metropolitan University, Hachioji 192-0397, Japan*
[3]*Center for Crystal Science and Technology, University of Yamanashi, Kofu 400-8511, Japan*

$Sn_{0.8}Ag_{0.2}Te$ is a new superconductor with $T_c$ ~ 2.4 K. The superconducting properties of $Sn_{0.8}Ag_{0.2}Te$ have been investigated by specific heat measurements under magnetic fields. Bulk nature of superconductivity was confirmed from the amplitude of the specific heat jump at the superconducting transition, and the amplitude is consistent with fully-gapped superconductivity. Upper critical field was estimated from specific heat and electrical resistivity measurements under magnetic fields. The Hall coefficient was positive, suggesting that the Ag acts as a p-type dopant in $Sn_{0.8}Ag_{0.2}Te$.

Recently, superconductors derived from a topological insulator [1-3] have been actively studied with the expecting of the emergence of Majorana fermions [4-8]. SnTe is one of the topological insulators (topological crystalline insulators) [9–11], and it becomes a superconductor by a partial substitution of Sn by In [12-16]. The superconducting transition temperature ($T_c$) of $Sn_{1-x}In_xTe$ increases with increasing In concentration and reaches 4.8 K for $x$ = 0.4–0.5. In the $Sn_{1-x}In_xTe$ superconductor, surface Andreev bound states, which was related to odd-parity pairing and topological superconductivity, were observed [17]. Therefore, SnTe-based superconductors are a potential candidate in which topological superconductivity could emerge.

Recently, we observed superconductivity with a $T_c$ ~ 2.4 K in Ag-substituted SnTe, $Sn_{1-x}Ag_xTe$ [18]. The $Sn_{1-x}Ag_xTe$ exhibits phase separations into the Ag-poor phase ($x$ ~ 0–0.1) and the $SnAgTe_2$ ($x$ ~ 0.5) phase when the samples were prepared at ambient pressure (by annealing in an evacuated quartz tube) [18,19]. However, the phase separation can be suppressed by high-pressure (HP) synthesis, and single-phase samples with a wide range of $x$ = 0–0.5 can be obtained using the HP-synthesis method. From the magnetic susceptibility measurements, the largest shielding volume fraction and the highest $T_c$ of 2.4 K were observed for $x$ = 0.2. Considering the structural similarity between $Sn_{1-x}In_xTe$ and

Sn$_{1-x}$Ag$_x$Te, we expect the emergence of topological superconductivity in Sn$_{1-x}$Ag$_x$Te as well. Therefore, the understanding of superconducting properties of Sn$_{1-x}$Ag$_x$Te is important for future investigations on superconductivity mechanisms. In this study, we performed specific heat and electrical resistivity measurements under magnetic fields for HP-Sn$_{0.8}$Ag$_{0.2}$Te (HP-synthesized Sn$_{0.8}$Ag$_{0.2}$Te). In addition, Hall measurements were performed to clarify the role of Ag substituted for Sn in Sn$_{0.8}$Ag$_{0.2}$Te.

The Sn$_{0.8}$Ag$_{0.2}$Te polycrystalline sample was prepared by HP synthesis as described in Ref. 18; in this study, we call this sample Sn$_{0.8}$Ag$_{0.2}$Te (or HP-Sn$_{0.8}$Ag$_{0.2}$Te). The Ag (99.9%), Sn (99.99%), and Te (99.9999%) powders with a nominal composition of Sn$_{0.8}$Ag$_{0.2}$Te were mixed and pressed into a pellet of 5 mm diameter. The pellet was heated at ~500 ºC for 30 min. under a pressure of 2 GPa. In addition, we compared specific heat data for HP-Sn$_{0.8}$Ag$_{0.2}$Te and AP-Sn$_{0.8}$Ag$_{0.2}$Te, which was prepared by heating the Sn$_{0.8}$Ag$_{0.2}$Te mixture in an evacuated quartz tube and did not show bulk superconductivity at above 1.8 K [18]. The lattice constant of the HP-Sn$_{0.8}$Ag$_{0.2}$Te and AP-Sn$_{0.8}$Ag$_{0.2}$Te samples used in this study was estimated as 6.1767 and 6.2365 Å, respectively, with a rock-salt cubic structure model (*Fm-3m*). Temperature dependence of specific heat (*C*) was measured by a relaxation method with Physical Property Measurement System (PPMS, Quantum Design) equipped with a $^3$He-probe system under magnetic fields of 0, 0.1, 0.25, 0.5, 0.75, 1.25, and 2 T. Specific heat measurements were also performed by a relaxation method using a dilution system under 0 T (the data in Fig. 4). Electrical resistivity was measured by a DC four-terminal method with the PPMS down to 0.5 K with a current of 300 mA (current density of 56 A/cm$^2$). Hall coefficient was measured by a four-terminal method using an AC resistivity mode (with a current of 20 mA) with a PPMS DynaCool (Quantum Design). The Hall measurements were performed at 4 and 300 K under a magnetic field of ±9 T.

Figure 1 shows the temperature (*T*) dependences of *C*/*T* of Sn$_{0.8}$Ag$_{0.2}$Te under magnetic fields of 0, 0.1, 0.25, 0.5, 0.75, 1.25, and 2 T. A sharp jump of *C*/*T*, corresponding to a superconducting transition, is observed at $T_c$ = 2.1 K. With increasing magnetic field ($\mu_0 H$), $T_c$ decreases, and superconductivity is almost suppressed at 0.75 T. In a previous study on magnetic susceptibility of Sn$_{0.8}$Ag$_{0.2}$Te, the upper critical field, $\mu_0 H_{c2}$ (0 K), was estimated as 0.72 T [18]. Thus, the suppression of superconductivity at above 0.75 T seems reasonable.

Figure 2 shows the *C*/*T*-*T*$^2$ plot of Sn$_{0.8}$Ag$_{0.2}$Te, measured at 2 T. Since the data points can be fitted with a linear function, we can estimate the electronic specific heat parameters ($\gamma$) as 2.95 mJK$^{-2}$mol$^{-1}$ because low-temperature specific heat can be described as $C/T = \gamma + \beta T^2$, where the $\beta$ parameter is the contribution of phonons. The obtained $\beta$ is 0.813 mJK$^{-4}$mol$^{-1}$. From the equation, $\beta = (12/5)\pi^4(2N)k_B\Theta_D^{-3}$, where *N*, $k_B$, and $\Theta_D$ are the Avogadro constant, the Boltzmann constant, and Debye temperature, $\Theta_D$, is calculated as 169 K. The estimated $\Theta_D$ for Sn$_{0.8}$Ag$_{0.2}$Te is closed to 162 K for $Sn_{0.6}In_{0.4}Te$ [16].

Using the estimated $\beta$ parameter, electron contribution of specific heat ($C_{el}$) is estimated by subtracting phonon contribution, which can be described as $\beta T^3$ at low temperatures. Figure 3 shows the $T$ dependences of $C_{el}/T$. The specific heat jump at $T_c$ ($\Delta C_{el}$) for $\mu_0 H = 0$ T is estimated as $\Delta C_{el} = 9.6$ mJ/K$^{-1}$mol$^{-1}$ using the solid line on the data at 0 T. From the obtained parameters, $T_c$, $\gamma$, and $\Delta C_{el}$, $\Delta C_{el}/\gamma T_c$ is calculated as 1.54, which is slightly larger than the value expected from the weak-coupling BCS approximation ($\Delta C_{SC}/\gamma T_c = 1.43$). The slightly larger superconducting gap may be resulting from strong-coupling nature, but it seems reasonable as a fully-gapped superconductor. As a result, bulk superconductivity of $Sn_{0.8}Ag_{0.2}Te$ has been confirmed. Since fully-gaped superconductivity does not contradict to the emergence of topological superconductivity in $Sn_{0.8}Ag_{0.2}Te$, further investigations on the superconductivity mechanisms of $Sn_{1-x}Ag_xTe$ are desired.

Next, we discuss the effect of HP synthesis to superconductivity of $Sn_{0.8}Ag_{0.2}Te$ from a specific heat viewpoint. As reported in Ref. 18, the $Sn_{0.8}Ag_{0.2}Te$ sample synthesized at ambient pressure (AP-$Sn_{0.8}Ag_{0.2}Te$) does not show superconductivity at $T > 1.8$ K. The lattice constant of AP-$Sn_{0.8}Ag_{0.2}Te$ is $a = 6.2365$ Å, which is clearly larger than $a = 6.1767$ Å of HP-$Sn_{0.8}Ag_{0.2}Te$. In $Sn_{1-x}Ag_xTe$, lattice constant basically decreases with increasing Ag concentration. Thus, the large lattice constant in AP-$Sn_{0.8}Ag_{0.2}Te$ would be due to phase separation due to solubility limit of Ag for Sn, while the X-ray diffraction patterns of both HP- and AP-$Sn_{0.8}Ag_{0.2}Te$ indicate almost single phase. Namely, we assume that local Ag-rich regions, which cannot be detected with laboratory XRD, coexist with the major phase of Ag-poor phase. Then, to get further information about the emergence of superconductivity in $Sn_{1-x}Ag_xTe$, comparison of specific heat of HP- and AP-$Sn_{0.8}Ag_{0.2}Te$ is important. Figure 4 shows the $T$ dependences of $C/T$ of HP-$Sn_{0.8}Ag_{0.2}Te$ and AP-$Sn_{0.8}Ag_{0.2}Te$ under a magnetic field of 0 T, which were measured using a dilution system. For HP-$Sn_{0.8}Ag_{0.2}Te$, the result is almost the same as that obtained in Fig. 1. For AP-$Sn_{0.8}Ag_{0.2}Te$, $C/T$ data at above 1.3 K almost correspond to the $C/T$ data for HP-$Sn_{0.8}Ag_{0.2}Te$, and a superconducting transition is observed at $T = 1.2$ K with a large specific heat jump. Thus, AP-$Sn_{0.8}Ag_{0.2}Te$ is also a bulk superconductor with a $T_c$ of 1.2 K. The difference of $T_c$ in HP- and AP-$Sn_{0.8}Ag_{0.2}Te$ is almost twice. One of the possible explanations of the difference is the phase separation scenario, mentioned above. The Ag-poor $Sn_{1-x}Ag_xTe$ phase is a major phase in AP-$Sn_{0.8}Ag_{0.2}Te$ and shows superconductivity with $T_c = 1.2$ K. Based on the $T_c$ of 1.2 K, actual Ag content in the superconducting phase is expected to be lower than $x = 0.2$. As will be mentioned later, the coherence length of this system is ~20 nm. Assuming that there are local Ag-rich phases with a size less than the coherence length in AP-$Sn_{0.8}Ag_{0.2}Te$, the observation of a large specific heat jump at 1.2 K, the emergence of bulk superconductivity, can be understood. This is also consistent with the fact that the phase separation of AP-$Sn_{0.8}Ag_{0.2}Te$ cannot be detected with laboratory XRD. On the basis of these results and discussion, we conclude that the merit of HP-synthesis should be the suppression of the phase separation into Ag-poor and Ag-rich phases, which achieves the optimal Ag concentration as a bulk superconductor.

$T_c$s of HP-Sn$_{0.8}$Ag$_{0.2}$Te determined from the specific heat measurements ($T_c^C$) are plotted in Fig. 5. For comparison, $T_c^{90\%}$ and $T_c^{zero}$ estimated from the temperature dependences of electrical resistivity ($\rho$) under magnetic fields up to 1 T, which are displayed in the inset of Fig. 5, are also plotted. The onset temperature of superconductivity in the $\rho$-$T$ measurements is defined as the temperature where $\rho$ becomes 90% of the normal state resistivity at just above $T_c$, as indicated with a dashed line in the inset figure. The $T_c^C$ and $T_c^{90\%}$ almost correspond to each other. From this phase diagram (Fig. 5), we note that the $T$ dependence of upper critical field ($\mu_0 H_{c2}$) is still almost linear at $T \sim 0.5$ K. Namely, the curve deviates from the Werthamer-Helfand-Hohemberg (WHH) model [20]. Here, Pauli paramagnetic effect does not affect because Pauli limiting field can be estimated as $1.84 \times T_c = 4.4$ T. Hence, from rough extrapolation of the data points, we estimate $\mu_0 H_{c2}$ (0 K) and irreversible field $\mu_0 H_{irr}$ (0 K) as 0.8‑0.9 T and 0.6‑0.7 T, respectively. Since upper critical field is described as $\mu_0 H_{c2} = \Phi_0/2\pi\xi^2$, where $\Phi_0$ and $\xi$ are quantum vortex ($h/2e$; $e$ is elementary charge) and coherence length, we can calculate $\xi$ as ~20 nm for HP-Sn$_{0.8}$Ag$_{0.2}$Te.

At the end, we briefly report the estimation of Hall coefficient ($R_H$) of HP-Sn$_{0.8}$Ag$_{0.2}$Te. In a recent study on Hall measurements and band structure calculations of Sn$_{1-x}$In$_x$Te, interesting In-doping dependence of $R_H$ and band structure evolution have been reported [16]. The $R_H$ for the In-poor phase is positive, which seems consistent with the assumption that the valence of Sn and Te is Sn$^{2+}$ and Te$^{2-}$, respectively, and In acts as a p-type dopant in SnTe. However, $R_H$ changes to negative at the In-rich phase. Therefore, the estimation of carriers in Sn$_{0.8}$Ag$_{0.2}$Te is important. The estimated $R_H$ of Sn$_{0.8}$Ag$_{0.2}$Te at 300 and 4 K is 0.0013 and 0.0043 cm$^3$/C, respectively. Band calculations in Ref. 16 indicate that rigid band shift occurs in the case of Ag substitution (3% and 12% substitutions) in Sn$_{1-x}$Ag$_x$Te. In addition, single-band approximation works in the analysis of carrier concentration in doped SnTe compounds [16]. Hence, we calculate carrier concentration ($n$) by assuming single-band approximation. $n$ for Sn$_{0.8}$Ag$_{0.2}$Te is calculated as $4.7 \times 10^{21}$ cm$^{-3}$ ($T = 300$ K) and $1.5 \times 10^{21}$ cm$^{-3}$ ($T = 4$ K) from the equation, $n = 1/eR_H$. The order of estimated $n$ roughly corresponds to the simulated value of $3.4 \times 10^{21}$ cm$^{-3}$, which is calculated on the basis of the refined lattice volume [18] and an assumption that Ag is Ag$^+$ in Sn$_{0.8}$Ag$_{0.2}$Te. These results suggest that the valence of Ag is basically +1 in Sn$_{0.8}$Ag$_{0.2}$Te, and Ag substitution generates p-type carriers in Sn$_{0.8}$Ag$_{0.2}$Te, which is the same situation with the In-poor phase of Sn$_{1-x}$In$_x$Te. However, in contrast to Sn$_{1-x}$In$_x$Te, superconductivity is suppressed with approaching Ag-rich region, namely in SnAgTe$_2$ [18]. To understand the emergence of superconductivity in SnTe-based compounds and to clarify the mechanisms of the superconductivity, further studies are needed.

In conclusion, we performed specific heat measurements for a HP-Sn$_{0.8}$Ag$_{0.2}$Te sample prepared by the HP-synthesis method. From a specific heat jump at $T_c$, bulk nature of the superconductivity of Sn$_{0.8}$Ag$_{0.2}$Te has been confirmed. $\Delta C_{el}/\gamma T_c$ for Sn$_{0.8}$Ag$_{0.2}$Te is 1.54, which is slightly larger than the value expected from the weak-coupling BCS model but seems reasonable as a fully-gapped superconductor. The temperature dependence of specific heat for AP-$Sn_{0.8}Ag_{0.2}Te$, which was synthesized at ambient pressure, was also investigated. We found that AP-$Sn_{0.8}Ag_{0.2}Te$ also showed bulk superconductivity with a $T_c$ of 1.2 K. From the specific heat and electrical resistivity measurements for HP-$Sn_{0.8}Ag_{0.2}Te$, a magnetic field-temperature phase diagram was established. The Hall measurements at 300 and 4 K proposed that the dominant carriers in $Sn_{0.8}Ag_{0.2}Te$ were holes.

**Acknowledgment**

This work was partly supported by Grants-in-Aid for Scientific Research (Nos. 15H05886, 15H05884, 25707031, and 16H04493).

*E-mail: mizugu@tmu.ac.jp

1) X. L. Qi and S. C. Zhang, Rev. Mod. Phys. **83**, 1057 (2011).
2) C. L. Kane, and E. J. Mele, Phys. Rev. Lett. **95**, 146802 (2005).
3) Y. L. Chen, J. G. Analytis, J. H. Chu, Z. K. Liu, S. K. Mo, X. L. Qi, H. J. Zhang, D. H. Lu, X. Dai, Z. Fang, S. C. Zhang, I. R. Fisher, Z. Hussain, and Z. X. She, Science **325**, 178 (2009).
4) L. Fu and C. L. Kane, Phys. Rev. Lett. **100**, 096407 (2008)
5) Y. Ando and L. Fu, Annu. Rev. Condens. Matter Phys. **6**, 361 (2015).
6) Y. S. Hor, A. J. Williams, J. G. Checkelsky, P. Roushan, J. Seo, Q. Xu, H. W. Zandbergen, A. Yazdani, N. P. Ong, and R. J. Cava, Phys. Rev. Lett. **104**, 057001 (2010).
7) S. Sasaki, M. Kriener, K. Segawa, K. Yada, Y. Tanaka, M. Sato, and Y. Ando, Phys. Rev. Lett. **107**, 217001 (2011).
8) M. Kriener, K. Segawa, Z. Ren, S. Sasaki, and Y. Ando, Phys. Rev. Lett. **106**, 127004 (2011).
9) L. Fu, Phys. Rev. Lett **106**, 106802 (2011).
10) T. H. Hsieh, H. Lin, J. Liu, W. Duan, A. Bansil, and L. Fu, Nat. Commun. **3**, 982 (2012).
11) Y. Tanaka, Z. Ren, T. Sato, K. Nakayama, S. Souma, T. Takahashi, K. Segawa, and Y. Ando, Nat. Phys. **8**, 800 (2012).
12) A. S. Erickson, J. H. Chu, M. F. Toney, T. H. Geballe, and I. R. Fisher, Phys. Rev. B **79**, 024520 (2009).
13) G. Balakrishnan, L. Bawden, S. Cavendish, and M. R. Lees, Phys. Rev. B **87**, 140507 (2013).
14) M. Novak, S. Sasaki, M. Kriener, K. Segawa, and Y. Ando, Phys. Rev. B **88**, 140502 (2013).
15) R. D. Zhong, J. A. Schneeloch, X. Y. Shi, Z. J. Xu, C. Zhang, J. M. Tranquada, Q. Li, and G. D. Gu, Phys. Rev. B **88**, 020505 (2013).
16) N. Haldolaarachchige, Q. Gibson, W. Xie, M. B. Nielsen, S. Kushwaha, and R. J. Cava, Phys. Rev. B **93**, 024520 (2016).

17) S. Sasaki, Z. Ren, A. A. Taskin, K. Segawa, L. Fu, and Y. Ando, Phys. Rev. Lett. **109**, 217004 (2012).
18) Y. Mizuguchi and O. Miura, J. Phys. Soc. Jpn. **85**, 053702 (2016).
19) M. P. Mathur, D. W. Deis, C. K. Jones, and W. J. Carr, Jr., J. Phys. Chem. Solids **34**, 183 (1973).
20) N. R. Werthamer, E. Helfand, and P. C. Hohemberg, Phys. Rev. **147**, 295 (1966).

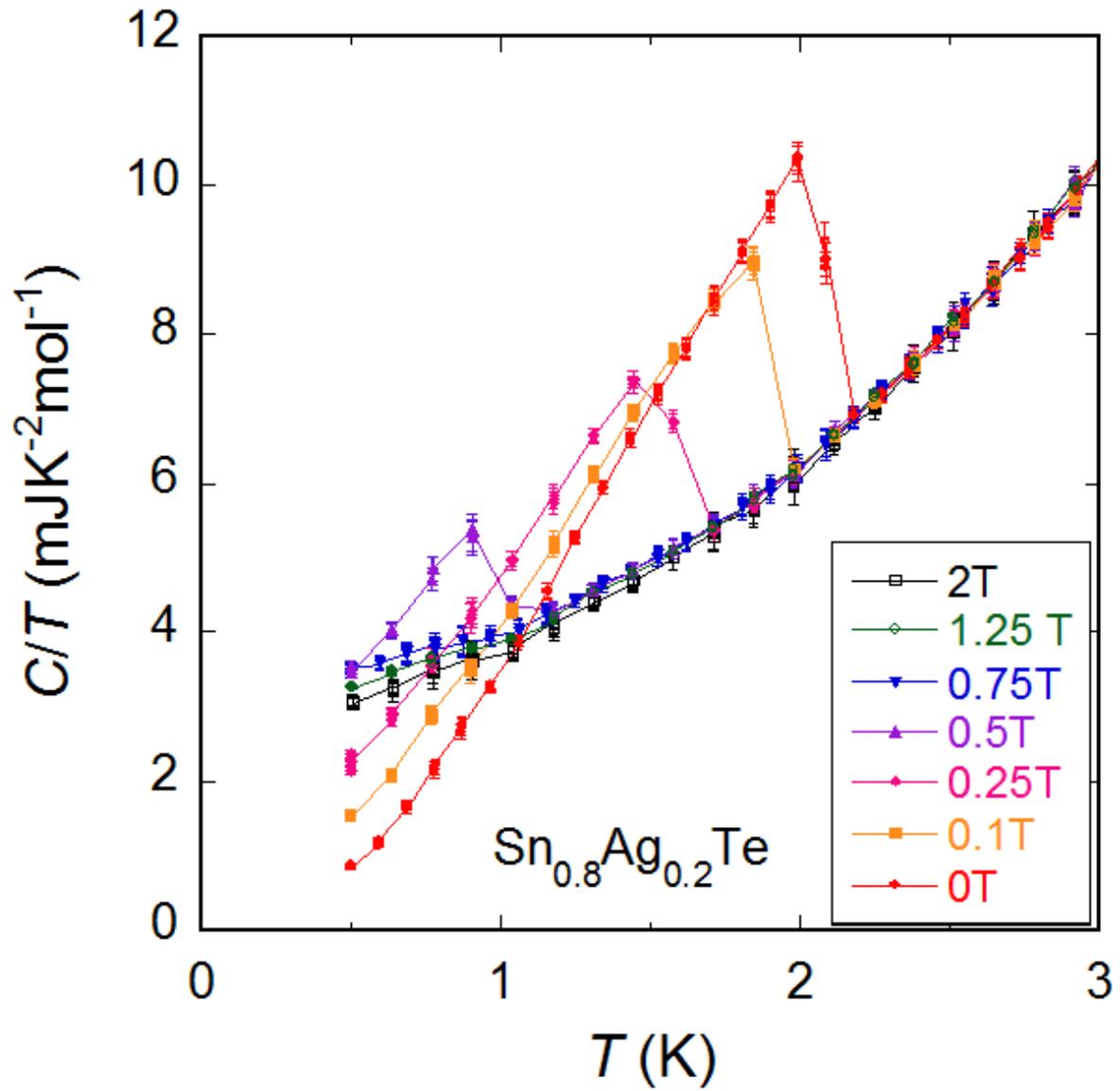

Fig. 1. Temperature (*T*) dependences of *C*/*T* of HP-Sn$_{0.8}$Ag$_{0.2}$Te under magnetic fields of 0, 0.1, 0.25, 0.5, 0.75, and 2 T.

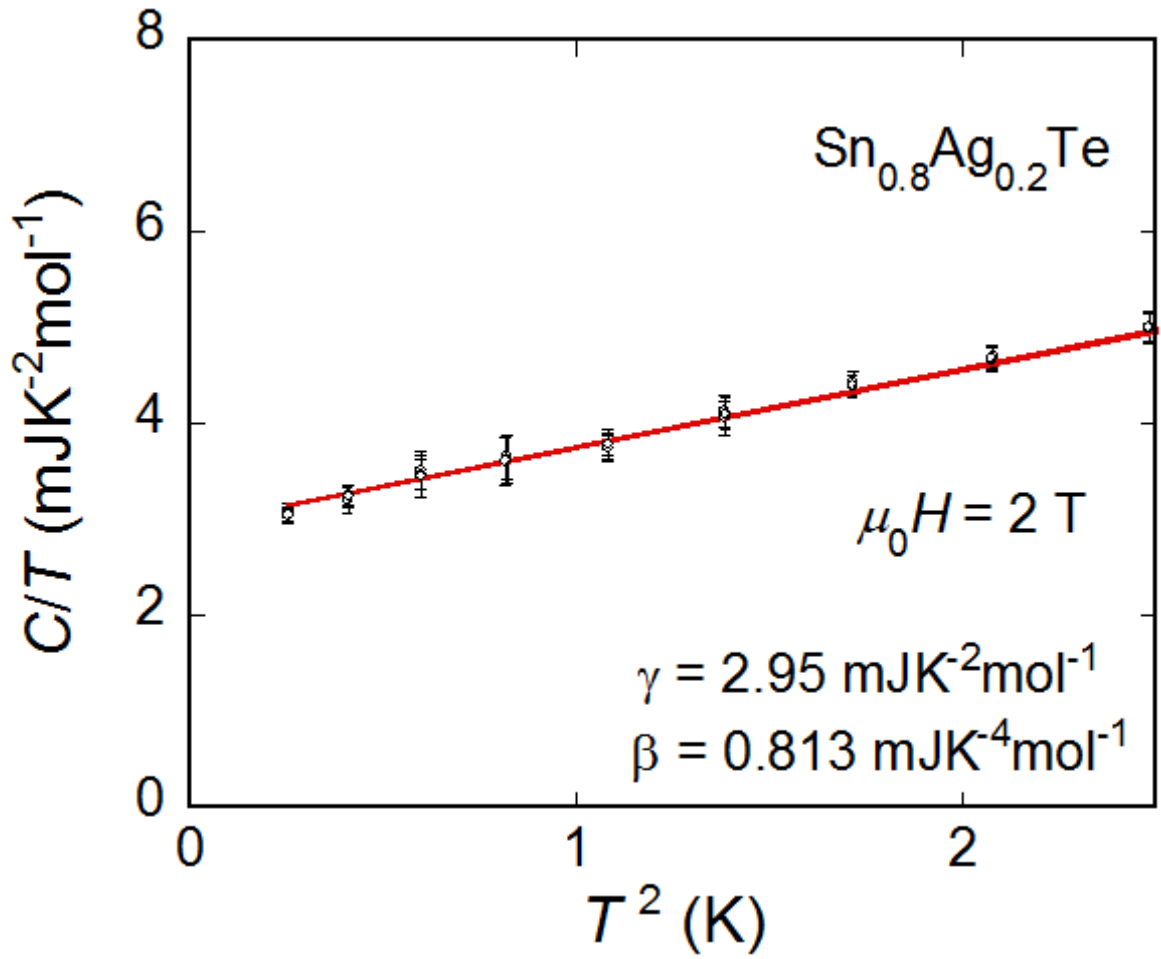

Fig. 2. $C/T$-$T^2$ plot of HP-$Sn_{0.8}Ag_{0.2}Te$ measured at 2 T.

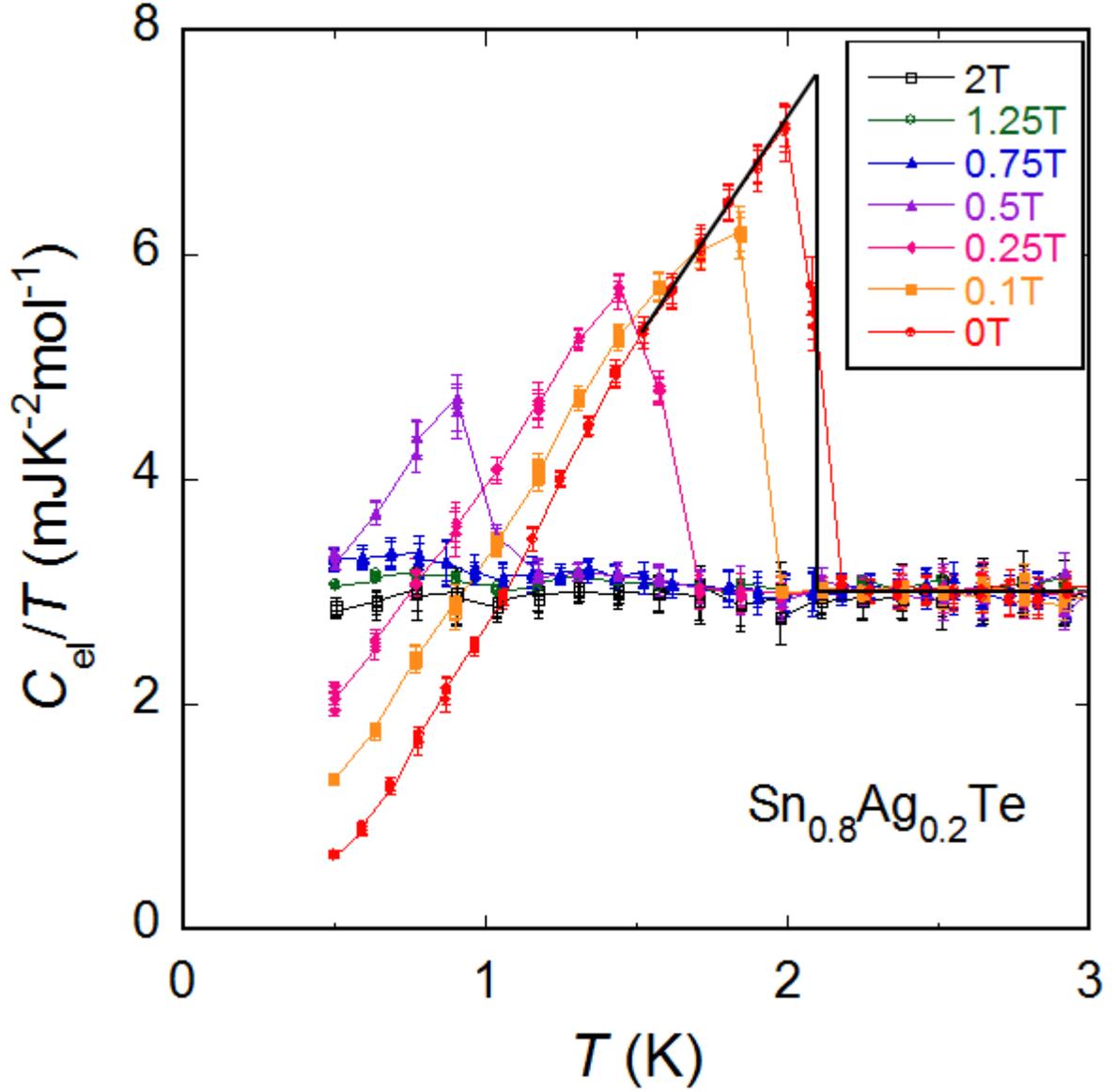

Fig. 3. $T$ dependence of $C_{el}/T$ of HP-$Sn_{0.8}Ag_{0.2}$Te under magnetic fields of 0, 0.1, 0.25, 0.5, 0.75, and 2 T. The solid line is used to estimate the specific heat jump ($\Delta C_{el}$) at $T_c$.

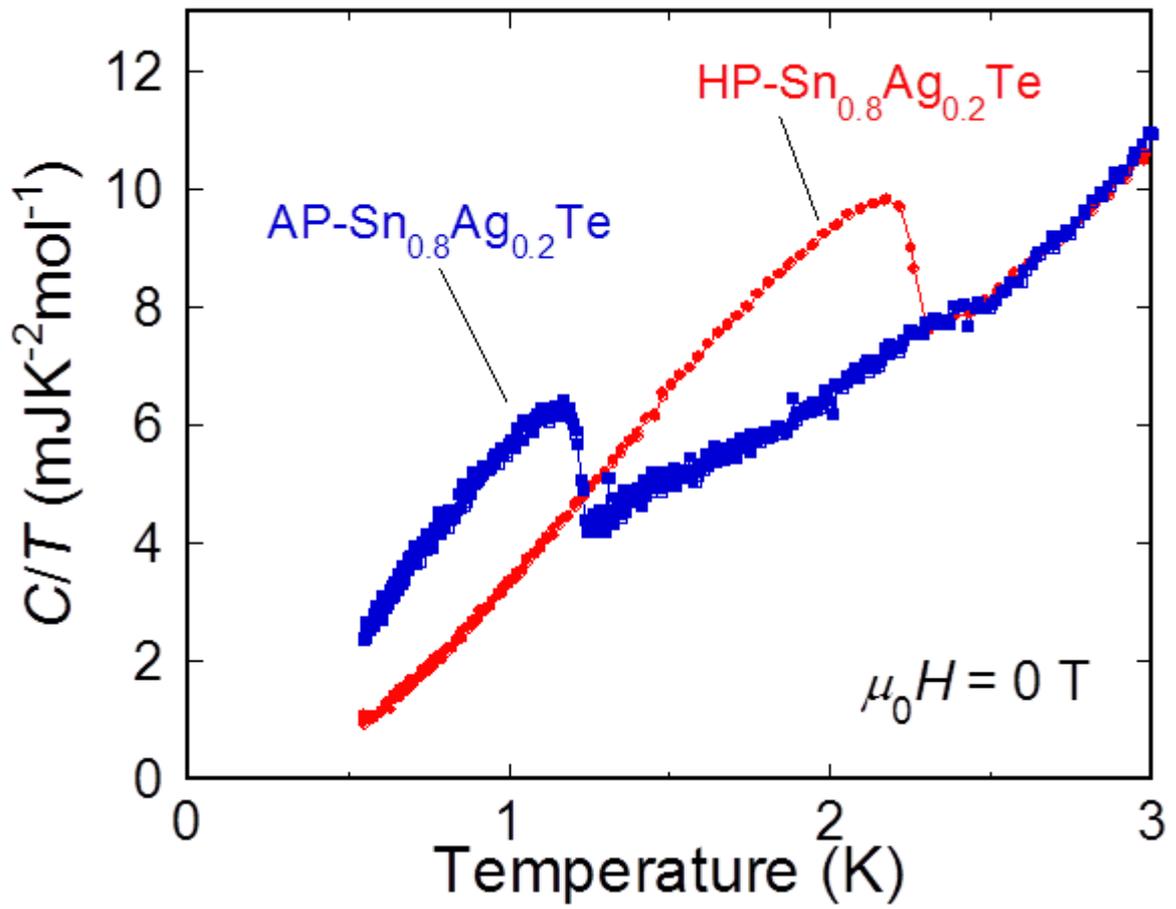

Fig. 4. Temperature dependences of C/T for HP-$Sn_{0.8}Ag_{0.2}Te$ and AP-$Sn_{0.8}Ag_{0.2}Te$.

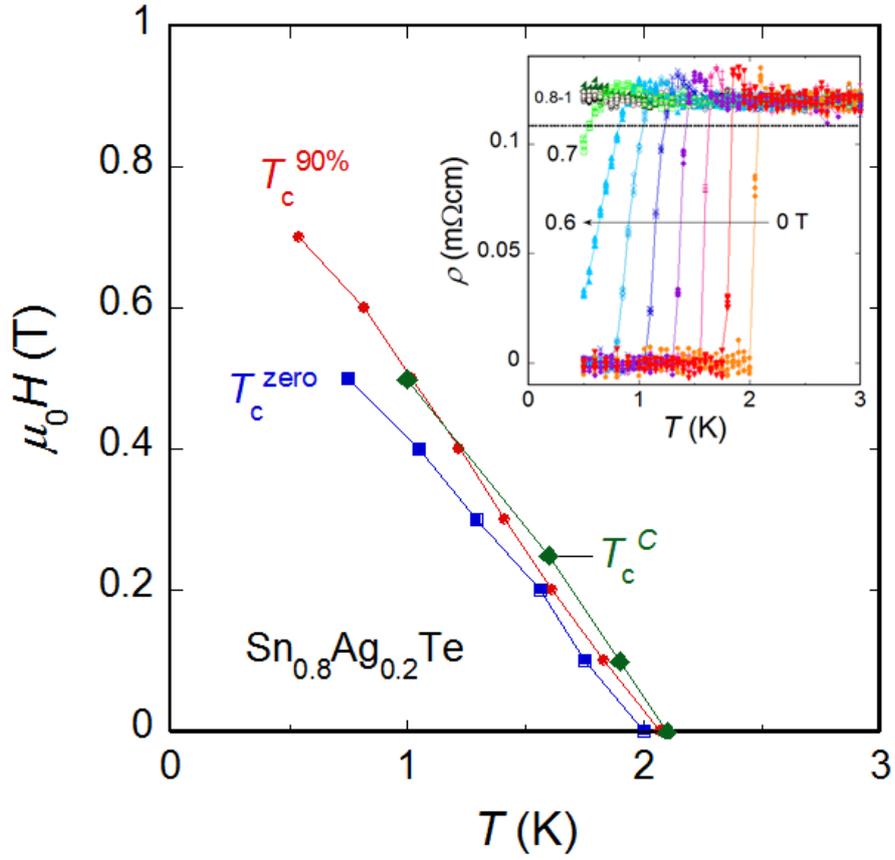

Fig. 5. Magnetic field-temperature phase diagram of HP-$Sn_{0.8}Ag_{0.2}Te$ with $T_c$s estimated from electrical resistivity measurements ($T_c^{90\%}$ and $T_c^{zero}$) and specific heat measurements ($T_c^C$). The inset shows the $T$ dependences of $\rho$ of HP-$Sn_{0.8}Ag_{0.2}Te$ under magnetic fields up to 1 T. We used 90% (dashed line) values of $\rho$ to estimate $H_{c2}$.